\documentclass[final,5p,times,twocolumn]{elsarticle}

\usepackage{graphicx}
\usepackage{amssymb}
\usepackage{bm}
\usepackage{amsmath}
\usepackage{color}

\begin{document}
\begin{frontmatter}

\title{Thin-shell theory based analysis of radially pressurized multiwall carbon nanotubes}

\author[1]{Hiroyuki Shima}
\author[2]{Susanta Ghosh}
\author[2]{Marino Arroyo}
\author[3]{Kohtaroh Iiboshi}
\author[3,4]{Motohiro Sato}

\address[1]{Division of Applied Physics,
Faculty of Engineering,
Hokkaido University, Sapporo, Hokkaido 060-8628, Japan}
\address[2]{Department of Applied Mathematics 3,
LaC\`aN, Universitat Polit\`ecnica de Catalunya, Barcelona 08034, Spain}
\address[3]{Division of Socio-Environmental Engineering, Faculty of Engineering,
Hokkaido University, Sapporo, Hokkaido 060-8628 Japan}
\address[4]{Department of Civil and Environmental Engineering,
Imperial College London, London SW7 2AZ, UK}

\begin{abstract}
The elastic radial deformation of multiwall carbon nanotubes (MWNTs) 
under hydrostatic pressure is investigated within the continuum elastic approximation.
The thin-shell theory, with accurate elastic constants and interwall couplings,
allows us to estimate the critical pressure 
above which the original circular cross-section transforms into radially corrugated ones.
The emphasis is placed on the rigorous formulation of the van der Waals interaction 
between adjacent walls, which we analyze using two different approaches.
Possible consequences of the radial corrugation in the physical properties of pressurized MWNTs
are also discussed. 
\end{abstract}


\end{frontmatter}

\section{Introduction}\label{sec1}

Nanomaterials are tiny platforms on which a beautiful interplay between
structure and property can be appreciated.
For the last decade, advanced nanofabrication techniques have put out
various nanostructured materials with novel geometry \cite{Tanda,Ozawa,Onoe,Tatewaki,Saranathan}, 
many of which exhibit unprecedented properties not seen 
in macroscopic structures. 
Among such nanostructures, carbon nanotubes have drawn great deal of attention. 
The salient feature of carbon nanotubes is their mechanical robustness and 
resilience \cite{ShimaPanPub}:
Due to the extremely large stiffness, for instance,
their thermal conductivity becomes 
higher than even that of diamond \cite{thermal}. 
Besides, their flexibility in bending \cite{Sears2004,Arias}, 
twisting \cite{Arias,XHuang,Zhang,Khademolhosseini},
radial compression \cite{Yakobson1996,Palaci2005,Gomez2006,Barboza,Diniz},
and the associated variations in the physical properties
hold promise for developing nanoelectromechanical devices \cite{LXDong,Roy}.
Towards successful implementations of such ideas,
computational studies have been playing 
a vital role in complementing experimental observations, often difficult and incomplete
for nanomaterials \cite{Rafii}.

In view of structure-property relations,
radial deformation is expected to elicit untouched behavior
of multiwall carbon nanotubes (MWNTs); see Fig.~\ref{fig_01} 
for a microscopic image of a typically synthesized MWNT \cite{Okita2007}.
The radial deformation disturbs both
the equal spacings between the concentric walls in MWNTs and 
the in-plane hexagonal lattice within each wall.
In particular, spatial modulation in the wall-wall separation
may enhance (or hinder) the electron charge transfer between neighboring walls
if they locally get close to (or pull away from) each other \cite{Zolyomi2008,Ahlskog2009};
this implies significant changes in various observables of MWNTs
under external pressure.
Despite the broad interest,
theoretical efforts on the mechanics of radially pressurized MWNTs remain 
limited \cite{ZXu2008,SHYang2009,XHuang2010}.
This scarcity is mainly because atomistic-based simulations
of MWNTs having many walls are computationally very expensive.

In the present work, we employ a powerful alternative,
the continuum elastic thin-shell theory approach,
to analyze the stable cross-sectional shapes of MWNTs under hydrostatic pressure.
Such approaches are often plagued with inaccurate or inconsistent elastic moduli. 
Here, we carefully select the intra-wall and wall-wall elastic constants for 
an accurate prediction of the critical pressure $p_c$ above which
the original circular cross-section transforms
into radially corrugated ones.
In the corrugation modes, each wall exhibits a wavy structure 
in the circumferential direction along the tube axis.
Such the pressure-induced corrugation is attributed to
the mismatch in the mechanical stability 
between the flexible outmost walls with large tube diameters
and rigid innermost walls with small diameters.

\section{Methods}

\subsection{Mechanical energy}

Although atomistic simulations may provide precise estimations
of physical quantities in general,
they often demand huge computational resources in systems of interest.
On this background, the thin shell theory based analysis
for the carbon nanotube mechanics
has long been developed \cite{Yakobson1996,Ru,Kudin_PRB2001,Pantano2004,YHuang2006,Chandraseker2007}. 
Along the continuum thin-shell method, a MWNT is mapped onto
a set of $N$ continuum elastic hollow tubes
of radii $r_i$ ($1\le i \le N$).
A point on the circle corresponding to the cross-section of the $i$th tube 
is described by
$(x,y) = (r_i\cos\theta, r_i\sin\theta)$
%
in terms of the polar coordinates; $\theta$ is the circumferential angle around the tube axis.
Under pressure $p$, the point moves to
\begin{eqnarray}
(x^*,y^*) &=&
\left( \;
\left[r_i+u_i(\theta, p)\right] \cos\theta - 
v_i (\theta, p) \sin \theta, \;\right. \\
& &
\left. \;
\left[r_i+u_i(\theta, p)\right] \sin\theta + 
v_i (\theta, p) \cos \theta \;
\right).
\label{eq_02}
\end{eqnarray}

\begin{figure}[ttt]
\vspace{0cm}
\begin{center}
\includegraphics[width=4.3cm]{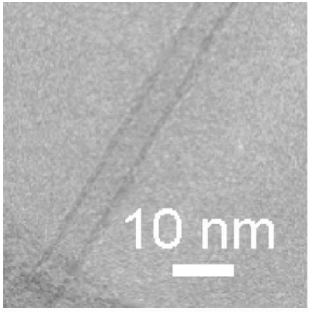}
\end{center}
\vspace{0cm}
\caption{Transmission electron microscopy image of 
a MWNT obtained via chemical vapor decomposition.
The number of concentric walls, $N$, and the innermost tube diameter, $D$,
are estimated as 
12 and 3.0 nm, respectively.
After Ref.~\cite{Okita2007}.}
\label{fig_01}
\end{figure}

If the deformation amplitudes, $u_i$ and $v_i$, 
are sufficiently small,
the mechanical energy of the $i$th tube per unit axial length
is given by
\begin{equation}
U_D^{(i)} = \frac{r_i}{2}
\left(
\frac{\tilde{C}}{1-\nu^2} \int_0^{2\pi} \epsilon_i^2 d\theta
+
\tilde{D} \int_0^{2\pi} \kappa_i^2 d\theta
\right).
\label{eq_03}
\end{equation}
Here, $\epsilon_i$ and $\kappa_i$ are respectively the
in-plane and bending strains of the $i$th wall,
both depending on $u_i$, $v_i$
and their derivatives with respect to $\theta$ \cite{Sanders}.
The constant $\tilde{C}$ denotes the in-plane stiffness,
$\tilde{D}$ the flexural rigidity,
and $\nu$ the Poisson ratio of each wall.

For quantitative discussions, the values of $\tilde{C}$ and $\tilde{D}$ must be carefully 
determined.
In the conventional thin-shell theory for macroscopic objects,
$\tilde{C}$ and $\tilde{D}$ are related to the Young's modulus $E$ 
of the wall and its thickness $\tilde{h}$ as
$\tilde{C} = E \tilde{h}$, $\tilde{D}=E \tilde{h}^3/[12 ( 1-\nu^2 )]$.
However, for carbon nanotubes, the wall is made out of 
a monoatomic 
graphitic layer and consequently the notion of a wall thickness becomes elusive. 
Hence, the macroscopic relations for $\tilde{C}$ and $\tilde{D}$ noted above
fail since there is no unique way
of defining the thickness of the graphene wall \cite{Gupta2010}.
Thus the values of $\tilde{C}$ and $\tilde{D}$ should be evaluated ab-initio
from direct measurements or computations of carbon sheets,
without reference to the macroscopic relations.
In actual calculations, we substitute $\tilde{C} = 345$ nN/nm,
$\tilde{D} = 0.238$ nN$\cdot$ nm, and $\nu =0.149$
along with the prior work \cite{Kudin_PRB2001} based on the density functional theory.
It should be noted that the values of $\tilde{C}$ and $\tilde{D}$ are
essentially tube-diameter dependent.
Neverthelss, such the dependences become negligible when the tube diameter exceeds 1 nm, 
above which the elastic constants of carbon nanotubes converge to those of a planar 
graphene sheet \cite{Kudin_PRB2001}.
On this background, we will take into account
only the nanotubes whose diameters are larger than 1 nm,
which allows to fix the values of $\tilde{C}$ and $\tilde{D}$ as noted above.

The stable cross-sections of MWNTs under $p$ 
minimize the mechanical energy $U$ of the whole system
that is described by \cite{NTN}
\begin{equation}
U = \sum_{i=1}^N U_D^{(i)} + \sum_{i,j=i\pm 1} U_I^{(i,j)} + \Omega,
\label{eq_01}
\end{equation}
where
\begin{equation}
U_I^{(i,j)} = 
\frac{c_{ij}(r_i+r_j)}{4} 
\int_0^{2\pi} (u_i - u_j)^2 d\theta
\label{eq_03}
\end{equation}
and
\begin{equation}
\Omega = p \int_0^{2\pi}
\left( r_N u_N + \frac{u_N^2 + v_N^2 - {u_N}' v_N + u_N {v_N}'}{2} \right) d\theta,
\label{eq_04}
\end{equation}
with $u' \equiv du/d\theta$.
The term $U_I^{(i,j)}$
accounts for the van der Waals (vdW) interaction energy of adjacent pairs of walls,
and $\Omega$ is
the negative of the work done by $p$ during cross-sectional deformation.
Note that $U$ is a functional of $u_i$ and $v_i$;
therefore, the variational method allows to obtain
the optimal solutions of $u_i$ and $v_i$ that minimize $U$ under a given $p$.

\begin{figure}[ttt]
\vspace{0cm}
\begin{center}
\includegraphics[width=4.5cm]{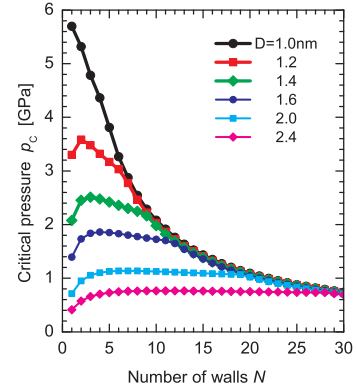}
\end{center}
\vspace{0cm}
\caption{(color online) $N$-dependence of critical pressure curve $p_c$.
Immediately above $p_c$, the original circular cross section of MWNTs gets
radially corrugated.}
\label{fig_03}
\end{figure}

For $N\gg 1$, the outside walls have large diameters, and consequently are very flexible and susceptible to mechanical instabilities under radial pressure. 
The contrast in the radial rigidity between the outermost walls
and the innermost ones
triggers a non-trivial cross-sectional deformation
observed in radially pressurized MWNTs.
In fact, 
the linearized buckling analysis leads
to the conclusion that
immediately above a critical pressure $p_c$, 
the circular cross section of MWNTs
becomes radially deformed as described by \cite{NTN}
\begin{equation}
u_i(\theta) = u_i^{(0)}(p_c) + \delta \mu_i(n) 
\cos n\theta,\;\;
v_i(\theta) = \delta \nu_i(n)
\sin n\theta.
\label{eq_05}
\end{equation}
The solution (\ref{eq_05}) represents a wavy structure
of a MWNT's cross-section,
called the radial corrugation with a mode index $n$.
The integer $n$ indicates the wave number of the corrugated walls,
being uniquely determined by 
the one-to-one relation between $n$ and $p_c$ \cite{NTN}.
We will find below that $n$ depends systematically on
$N$ and the innermost tube diameter $D \equiv r_1$ 
as presented in Fig.~\ref{fig_03}.

\subsection{Wall-wall interaction coefficient}

It is known that in the axial buckling of MWNTs, 
the buckling load is sensitive to the vdW interaction 
between adjacent walls \cite{He2005JMPS}.
Likewise, $p_c$ 
is thought to depend on the strength of vdW interactions;
this is the reason why we address the rigorous formulation of the interaction.
Using both a \emph{discrete-pairwise} summation and a \emph{continuum} approach, 
we derive the coefficients $c_{ij}$ in Eq.~(\ref{eq_03}) through 
a first order Taylor approximation of the vdW pressure \cite{He2005JMPS,WBLu_APL2009}
associated with the vdW potential
$V(d) = 4\varepsilon [(\sigma/d)^{12} - (\sigma/d)^6 ]$.
Here, $d$ is the distance between a pair of carbon atoms, $2^{1/6} \sigma = 0.3833 \,\mathrm{nm}$
is the equilibrium distance between two interacting atoms,  and
$\varepsilon=2.39 \,\mathrm{meV}$ is the well depth of the potential \cite{Girifalco_PRB2000}.
The resulting equilibrium spacing between neighboring walls, used here to define the geometry of the MWNCTs, is $0.3415 \,\mathrm{nm}$.
The derivative $F=\partial V/\partial d$  is the force between the two atoms.

There exist several continuum models for the vdW interactions 
\cite{He2005JMPS,WBLu_APL2009,Girifalco_PRB2000}.
In Ref.~\cite{He2005JMPS}, expressions for the pressure 
and the vdW interaction coefficients were obtained by integrating 
the continuum vdW force and its derivative on curved wall surface, 
while disregarding the vectorial nature of the force. 
The significance of the vectorial nature of 
the force was addressed in Ref.~\cite{WBLu_APL2009}, 
where analytical expressions for the vdW pressure were obtained
by considering only the component of the vdW force normal to the wall.
It was emphasized in Ref.~\cite{WBLu_APL2009}
that for a two-walled carbon 
nanotube,
the pressure exerted on inner wall is different 
from
the pressure on the outer wall. 
The pressures on the inner and outer walls for a concentric two-walled tube with radii 
$r_{\rm inn}$ and $r_{\rm out}$ are given below 
(with positive signs for compression):
\begin{equation}\label{eq_formula}
p_{\rm inn} = \alpha\frac{r_{\rm out}}{r_{\rm inn}}\,f_{-}
\quad {\rm and} \quad
p_{\rm out} =  \alpha\frac{r_{\rm inn}}{r_{\rm out}}\,f_{+},
\end{equation}
where
$\alpha=3\pi\varepsilon \sigma {\rho_{c}}^{2}/32$ with
$\rho_{c} = 38.18$ nm$^{-2}$ being the area density of carbon atoms.
We used the notations:
\begin{equation}
f_{\pm} =
231 \beta^{11}\left(\gamma\,E_{13} \pm E_{11}\right)
-160\beta^5\left(\gamma\,E_{7} \pm E_{5} \right),
\label{eq_fpm}
\end{equation}
where $\beta=\sigma/(r_{\rm out}+r_{\rm inn})$, $\gamma=h/(r_{\rm out}+r_{\rm inn})$,
$h=r_{\rm out}-r_{\rm inn}$,
$E_{m}=\int_{0}^{\pi/2}(1-k^{2}{\sin}^{2}\theta)^{-m/2}d\theta$ and
$k=4r_{\rm inn}r_{\rm out}/(r_{\rm inn}+r_{\rm out})^{2}$.

\begin{figure}[ttt]
\vspace{0cm}
\begin{minipage}{7.8cm}
\begin{center}
\includegraphics[width=5.8cm]{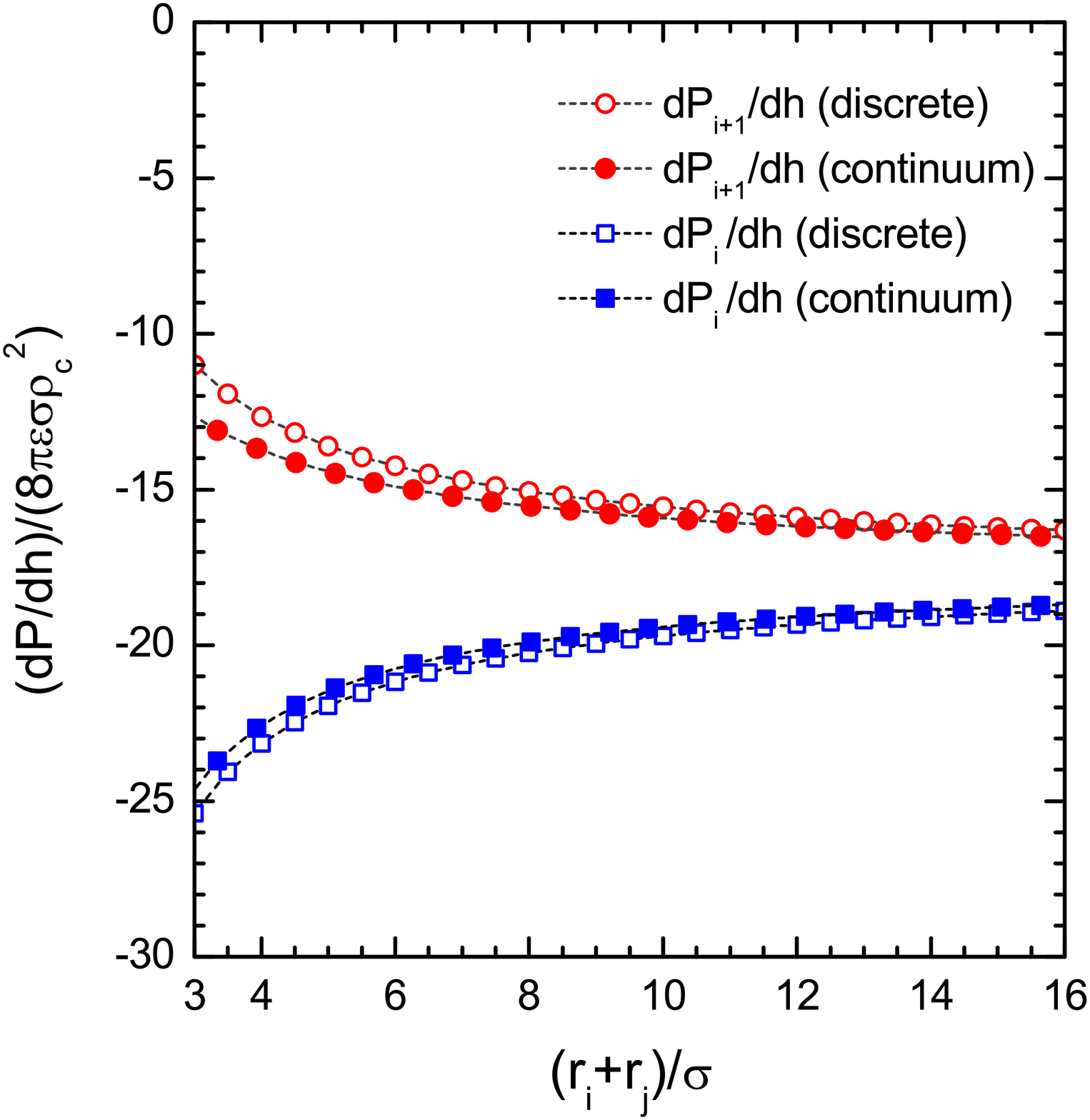}
\end{center}
\end{minipage}
\vspace{0cm}
\begin{minipage}{8cm}
\begin{center}
\includegraphics[width=5.7cm]{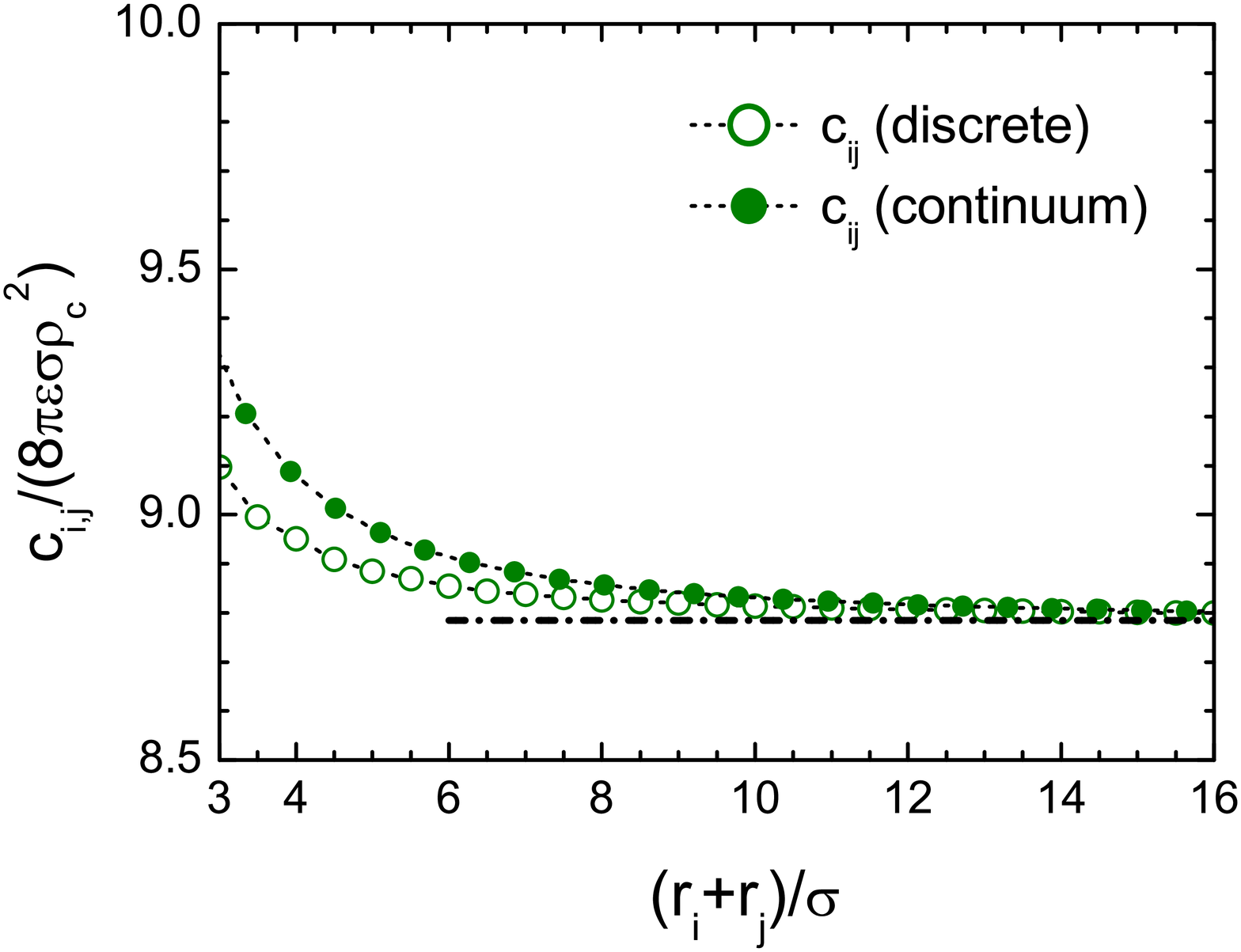}
\end{center}
\end{minipage}
\vspace{0cm}
\caption{Tube-radius dependence of the derivatives of pressure (top) and the vdW interaction coefficients $c_{i,j}$ (bottom).}
\label{fig_cij}
\end{figure}

In the following, we obtain analytical expressions for $c_{i,j}$ 
linearizing the formula (\ref{eq_formula}) for the pressure.
The formula takes into account correctly 
the normal-to-wall component of vdW forces, and avoids
common assumptions like $p_{\rm inn}=p_{\rm out}$ \cite{Pantano2004} and 
$p_{\rm inn}r_{\rm inn}=p_{\rm out}r_{\rm out}$ \cite{Ru2001JMPS};
therefore, the resulting $c_{i,j}$ will be also free from unnecessary assumptions.

In the present work, infinitesimal deformation is considered. 
Hence, the linearzed pressure is needed for vdW energy calculation. 
Note that the vdW energy depends quadratically on the change in spacing between two adjacent walls. 
Consider two consecutive walls with radii $r_{i}$ and $r_{i+1}$, 
where the subscripts $i$ and $i+1$ correspond to $inn$ and $out$ respectively. 
The vdW energy stored due to symmetric perturbation 
along the positive direction of pressure is given by
\begin{eqnarray}\label{eq_stored}
U \approx \frac{r_{\rm m}}{2} \int_{0}^{2\pi} 
\left(
-p_{i,i+1}\frac{\Delta h}{2}-p_{i,i+1}\frac{\Delta h}{2}
\right) d\theta,
\end{eqnarray}
where $r_{\rm m}=(r_{i}+r_{i+1})/2$ is the mean radius 
and $p_{i,i+1}$ is the vdW pressures on the $i$th wall. 
The corresponding linearized pressure is given by 
$\left.{\partial{p_{i,i+1}}}/{\partial h}\right| _{r_{\rm m}}$. 
In Eq.~(\ref{eq_stored}), $r_{\rm m}d\theta$ describes
the length of the infinitesimal element  on which the pressure is acting.
Using the linearized pressure and comparing with Eq.~(\ref{eq_03}),
we get the expressions for vdW coefficients as
\begin{eqnarray}
c_{i,i+1} = -\frac14 
\left(
\frac{\partial{p_{i,i+1}}}{\partial h}
+
\frac{\partial{p_{i+1,i}}}{\partial h}
\right),
\end{eqnarray}
where
\begin{eqnarray}\label{equ:pressure_coeff}
\left. \dfrac{\partial p_{\rm i,i+1}}{\partial h}\right| _{r_{\rm m}}
= \dfrac{2\,\alpha\,r_{\rm m}}{\left(r_{\rm m}-h/2\right)^{2}}\,f_{-} \,+\, \alpha
  \left(\frac{r_{\rm m}+h/2}{r_{\rm m}-h/2}\right)\,\dfrac{\partial f_{-}}{\partial h}, \nonumber \\
\left. \dfrac{\partial p_{\rm i+1,i}}{\partial h}\right| _{r_{\rm m}}
= -\dfrac{2\,\alpha\,r_{\rm m}}{\left(r_{\rm m}+h/2\right)^{2}}\,f_{+} \,+\, \alpha
\left(\frac{r_{\rm m}-h/2}{r_{\rm m}+h/2}\right)\,\dfrac{\partial f_{+}}{\partial h}.
\end{eqnarray}
See Appendix for the derivatives of $\partial f_{\pm}/\partial h$. 
Note that the $c_{i,j}$ is symmetric. 

These analytical expressions for the stiffness coefficients associated with 
the vdW interactions obtained with the continuum approximation are compared 
with the coefficients obtained using a discrete pairwise calculation. 
In the later approach, the pressure coefficients are computed by summing 
the component normal to the wall of the linearized vdW forces acting on an atom $a$,
\emph{i.e.} the normal component of 
$\sum_b (\partial F/\partial d) (\partial d/\partial h)|_{(d_{a-b})}$, 
where the linearized force $\partial F/\partial d$ is evaluated at 
the actual distance between the pair of interacting atoms $a$ and $b$, $d_{a-b}$.
In Ref.~\cite{He2005JMPS}, the linearization is done at the equilibrium distance of the potential.
Noting that most of the interacting atom pairs are not at the equilibrium distance, it is apparent that
the linearization should be done at the actual distance between each pair of interacting atoms.
The discrete pressure can be obtained by dividing the force exerted on one atom by the area per atom. 
The vdW force on one atom is obtained by considering the interactions 
between this atom with all the atoms on a neighboring wall.
The effect of the relative lattice placement
of interacting walls is averaged out 
by integrating the interactions over a symmetric triangle of the representative hexagonal cell of graphene.

The  normalized pressure derivatives 
$(\partial p_{i,j} /\partial h)/(8\pi\varepsilon\sigma{\rho_{c}}^{2})$ 
and normalized vdW interaction coefficients
$c_{i,j}/(8\pi\varepsilon\sigma{\rho_{c}}^{2})$ are 
plotted against normalized radius $(r_{\rm inn}+r_{\rm out})/\sigma$ in 
Fig.~\ref{fig_cij}.
A good match is observed between the \emph{discrete-pairwise} and the \emph{continuum} approach.
The interaction coefficient for two horizontally placed graphene walls, 
$c_{i,j}^{\rm graphene}$, 
is obtained from the second derivative of the analytic continuum interaction 
potential \cite{Girifalco_PRB2000}.
For larger radii,
$c_{i,j}$ approaches to $c_{i,j}^{\rm graphene}$, 
corroborating our results for nanotubes.

\section{Results and discussion}

\begin{figure}[ttt]
\vspace{0cm}
\begin{center}
\includegraphics[width=5.0cm]{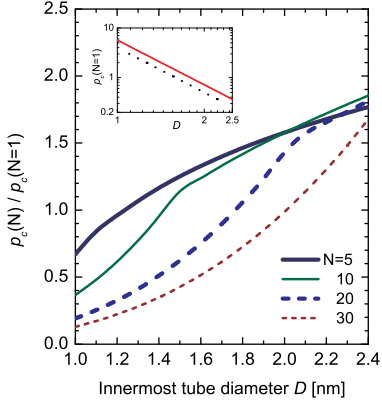}
\end{center}
\vspace{0cm}
\caption{(color online) $D$-dependence of 
normalized critical pressure curve $p_c$.
Inset: Cubic behavior of the $p_c$ curve when $N=1$.
The dashed line showing the power-law $p_c \propto D^{-3}$ is a visual reference.}
\label{fig_04}
\end{figure}

Figure \ref{fig_03} plots $p_c$ as a function of $N$ for various values of $D$.
An initial increase in $p_c$ at small $N$ (except for $D=1.0$ nm) 
is attributed to the enhancement of radial stiffness of the entire MWNT by encapsulation.
This stiffening effect disappears with further increase in $N$,
resulting in decay or convergence of $p_c(N)$.
It is noteworthy that MWNTs practically synthesized often show $D$
larger than those presented in Fig.~\ref{fig_03}.
In fact, the MWNT depicted in the image of Fig.~\ref{fig_01}
gives $D = 3.0$ nm, for which $p_c(N)$ lies at 
several hundreds of MPa as estimated from Fig.~\ref{fig_03}.
Such degree of pressure applied to MWNTs
is easily accessible in high-pressure experiments,
supporting the feasibility of our theoretical predictions.
The $D$-dependence of $p_c$ for various fixed $N$
is provided in Fig.~\ref{fig_04}
in a normalized manner;
When $N=1$, we obtain $p_c \propto D^{-3}$ in agreement with previous studies 
on single-walled nanotubes \cite{Yakobson1996,DYSun2004}.
It also deserves comment that a radial pressure
large enough to cause corrugation can be achieved by electron-beam irradiation;
the self-healing nature of eroded carbon walls gives rise to
a spontaneous contraction that exerts a high pressure on the inner walls
\cite{Banhart1996,LSun2006,Shima2010}

\begin{figure}[ttt]
\vspace{0cm}
\begin{center}
\includegraphics[width=4.7cm]{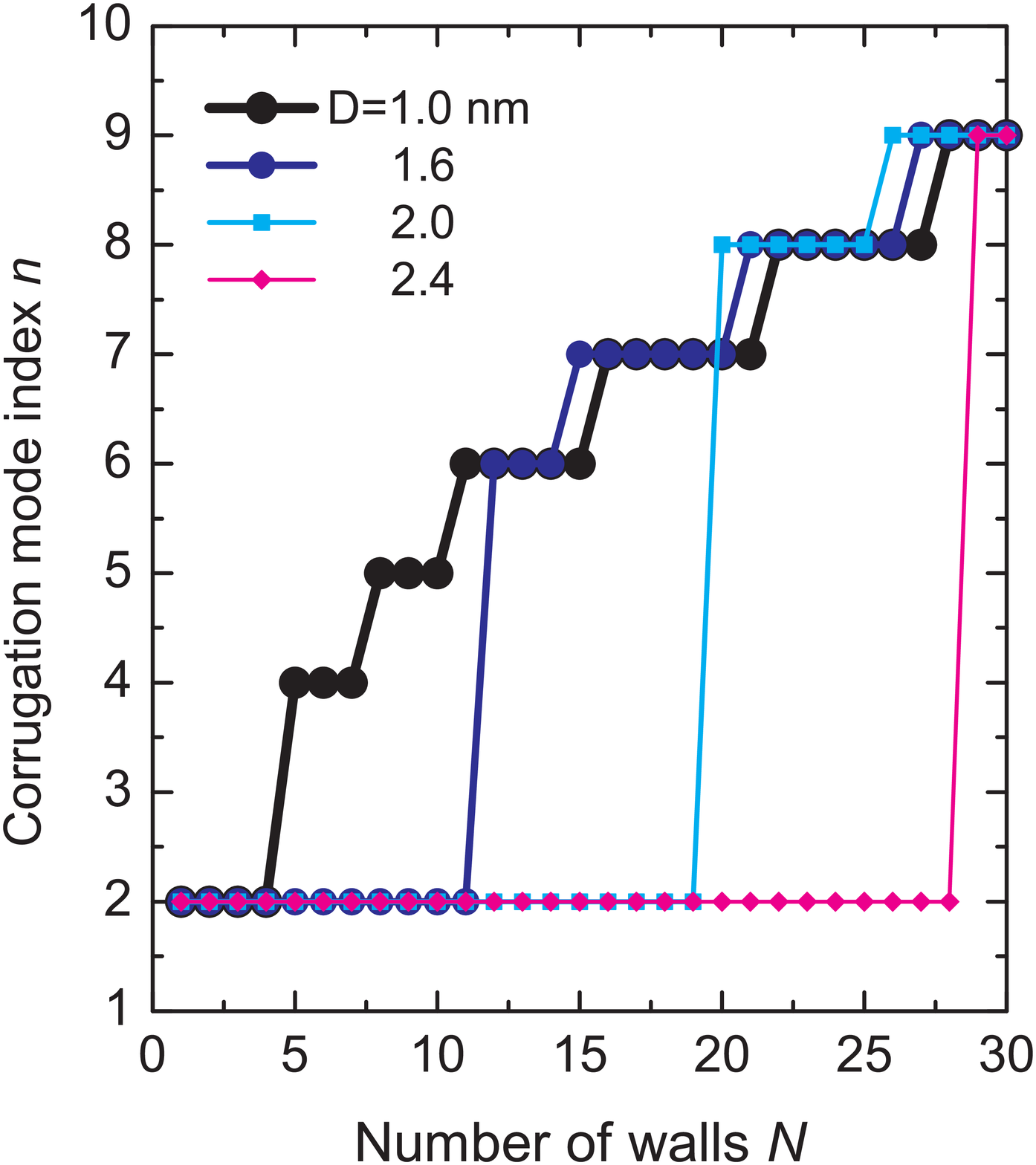}
\end{center}
\vspace{0cm}
\caption{(color online) Stepwise increase in the corrugation mode index $n$.}
\label{fig_05}
\end{figure}

Figure \ref{fig_05} shows the stepwise increases in the corrugation mode index $n$,
defined by Eq.~(\ref{eq_05}).
For all $D$, the deformation mode observed just above $p_c$ abruptly
increases from $n=2$ to $n\ge 4$ at a certain value of $N$,
followed by the successive emergence of higher corrugation modes with larger $n$.
These successive transitions in $n$ at $N\gg 1$
originate from the mismatch in the radial stiffness of the innermost and outermost walls.
A large discrepancy in the radial stiffness of the inner and outer walls
results in a maldistribution of the deformation amplitudes
of concentric walls interacting via vdW forces,
which consequently produces an abrupt change in the observed deformation mode at a certain value of $N$.

A possible physical consequence of radial corrugation
is a pressure-driven change in quantum transport
of carriers in radially corrugated nanotubes.
It is known that mobile carriers whose motion is confined to
a thin curved layer
behave differently from those on a conventional flat plane
because of the curvature-induced electromagnetic field \cite{ShimaTLL,Ono,Taira}.
Another interesting issue is the effect of atomic lattice registry
on non-linear deformation ({\it i.e.} large deformation amplitude) 
regimes \cite{Yu_regist,Liu_regist,Marom_regist}.
In the latter case,
the degree of commensurance in atomic structures between neighboring carbon layers
plays a prominent role
in determining the optimal morphology of highly corrugated MWNTs,
in which the interwall spacings partially vanish.
Crumpling or twisting caused by radial pressure is expected, 
and in fact were observed in preliminary calculations \cite{Mohammad}.

\section{Conclusion}

The thin-shell theory has been employed to derive
the critical hydrostatic pressure $p_c$ 
above which the cross section of pressurized MWNTs transforms into
radially corrugated patterns.
The value of $p_c$ lies on the order of several hundreds of MPa 
for practically synthesized MWNTs with $N\gg 1$ and $D\ge 3$ nm, 
indicating the experimental feasibility of the prediction.
The enegetically favored corrugation pattern strongly depends on $N$ and $D$,
where a larger $N$ and smaller $D$ give a larger corrugation mode index $n$.
Our results provide a clue in developing MWNT-based devices 
operating under high pressure.

\section*{Acknowledgement}

We acknowledge M.~Rahimi, E.~Osawa, Y.~Suda, K.~Yakubo and T.~Mikami
for helpful discussions.
HS is thankful for the financial supports provided by KAKENHI and Hokkaido Gas Co., Ltd.
SG acknowledges the support of the Spanish Ministry of Science and
Innovation through the Juan de la Cierva program.
MA acknowledges the support of the European Research Council
(FP7/2007-2013)/ERC grant agreement nr 240487 and the prize
``ICREA Academia'' funded by the Generalitat de Catalunya.
Numerical simulations were partly carried out using
the facilities of the Supercomputer Center, University of Tokyo.

\appendix

\section{Functional forms of $\partial f_{\pm}/\partial h$}\label{tba}

It follows from Eq.~(\ref{eq_fpm}) that
\begin{equation}
\begin{array}{l}
\left. \dfrac{\partial f_{\pm}}{\partial h} \right|_{r_{\rm m}}
= 
231\,\beta^{11} \left( {\frac {d\gamma}{dh}} {E_{13}}+\gamma{\frac {d{E_{13}}}{dh}}\pm{\frac {d{E_{11}}}{dh}} \right)  \\
\qquad \,   
-160\,\beta^{5} \left(  {\frac {d\gamma}{dh}} {E_{7}}+\gamma{\frac {d{E_{7}}}{dh}}\pm{\frac {d{E_{5}}}{dh}}\right). 
\label{eq_appdx}
\end{array} 
\end{equation}
All $E_{m}$ functions obey the recurrence relation
$(m-2)(1-k^{2})E_{m}=(m-3)(2-K^{2})E_{m-2}-(m-4)E_{m-4}$
from complete elliptic integrals
of the first and second kind, $K(k)$ and $E(k)$, respectively;
$E_{1}=K(k)$ and $E_{-1}=E(k)$ in the current notation. 
Hence, the required derivatives of $E_m$ in Eq.~(\ref{eq_appdx})
are given below:
\begin{equation}
\begin{array}{l} 
\frac{\partial E_{5}}{\partial k}
=
\frac{1}{3 \left( k^2-1 \right)^3 k}
\left( - 3 E -7 k^2 E + 2 k^4 E  + 3 K - 2 k^2 K - k^4 K \right), \\
\frac{\partial E_{7}}{\partial k}
=
\frac{1}{15\left(k^2-1\right)^4 k} 
\left(
15\, E +58\, k^2 E -33\, k^4 E + 8\, k^6 E
\right.\\ 
\qquad \qquad \qquad 
\left.
-15\, K + 2\, k^2 K + + 17\, k^4 K -4\, k^6 K
\right), \\ 
\frac{\partial E_{11}}{\partial k}
=
-\frac {1}{315 \left( -1+k^2 \right)^6 k} 
\left(
- 315\, k E -2309\, k^2 E + 2611\, k^4 E 
\right. \\ 
\qquad \qquad 
-1899\, k^6 E + 760\, k^8 E -128\, k^{10} E \\
\qquad \qquad
+315\, k K  + 419\, k^2 K - 1403\, k^4 K  + 993\, k^6 K  \\ 
\qquad \qquad 
\left.
- 388\, k^8 K  + 64\, k^{10} K \right), \\
\frac{\partial E_{13}}{\partial k}
=
\frac{1}{3465 \left( -1+k^2 \right)^7 k }
\left( -3465\,E -31907\, k^2 E +44991\,k^4 E     \right.\\  
\qquad \quad - 43633 k^6 E + 26206\,k^8 E - 8832\,k^{10} E +1280\, k^{12} E  \\  
\qquad \quad +3465\, K +7652\, k^2 K -24562\,k^4 K  +23204\,k^6 K  \\  
\qquad \quad  \left.-13615\,k^8 K  + 4496\,k^{10} K -640\, k^{12} K \right).
\end{array}
\end{equation}
Other derivatives are given by
$\partial k/\partial h|_{r_m} = -h/(r_m^2 k)$ and $\partial \gamma/\partial h|_{r_m}=1/r_m$.


\end{document}